\documentclass[12pt]{iopart}

\usepackage[english]{babel}
\usepackage{hyperref}
\hypersetup{
colorlinks=true,  
citecolor=black,
linkcolor=black,
}

\usepackage{graphicx}
\usepackage{ulem}
\usepackage{bbm}

\newcommand{\think}[1]{{\color{black} #1}}

\newcommand{\be}{\begin{equation}}
\newcommand{\beq}{\begin{eqnarray}}
\newcommand{\ee}{\end{equation}}
\newcommand{\eeq}{\end{eqnarray}}

\newcommand {\apgt}{\ {\raise-.5ex\hbox{$\buildrel<\over\sim$}}\ }
\newcommand {\aplt}{\ {\raise-.5ex\hbox{$\buildrel>\over\sim$}}\ }

\begin{document}

\title{Phase-noise protection in quantum-enhanced differential interferometry}

\author{M. Landini$^{1,2}$, M. Fattori$^{1,2,3}$, L. Pezz\`e$^{1,2,4}$ and A. Smerzi$^{1,2,4}$}
\address{$^1$Istituto Nazionale di Ottica-CNR (INO-CNR), Via Nello Carrara 1, I-50019 Sesto Fiorentino, Italy}
\address{$^2$European Laboratory for Non-Linear Spectroscopy (LENS) and Dipartimento di Fisica, Universit\'a di Firenze, Via Nello Carrara 1, I-50019 Sesto Fiorentino, Italy} 
\address{$^3$Istituto Nazionale di Fisica Nucleare (INFN), Sezione di Firenze, Via Sansone 1, I-50019 Sesto Fiorentino, Italy}
\address{$^4$QSTAR, Largo Enrico Fermi 2, 50125 Firenze, Italy} 

\ead{luca.pezze@ino.it}
\begin{abstract}
Differential interferometry (DI) with two coupled sensors is a most powerful approach for precision measurements in presence of
strong phase noise.
However DI has been studied and implemented only with classical resources.
Here we generalize the theory of differential interferometry to the case of entangled probe states.
We demonstrate that, \think{for perfectly correlated interferometers and} in the presence of arbitrary large phase noise, sub-shot noise sensitivities -- up to the Heisenberg limit -- 
are still possible with a special class of entangled states \think{in the ideal lossless scenario}. 
These states belong to a decoherence free subspace where entanglement is passively protected.
Our work pave the way to the full exploitation of entanglement in precision measurements in presence of strong phase noise.
\end{abstract}



\maketitle

\section{Introduction}
Atom interferometers~\cite{CroninRMP2009} offer nowadays unprecedented precision in the measurement of gravity~\cite{AchimNATURE1999}, 
inertial forces~\cite{GustavsonPRL1997}, atomic properties~\cite{EkstromPRA1995} and fundamental constants~\cite{BouchendiraPRL2011}. 
Their large sensitivity makes almost unavoidable their coupling to the environment
which mainly results in a random noise which affects the signal phase. 
In order to overcome this limitation, many experiments aiming at 
precision measurements adopt a differential scheme:
two interferometers operating in parallel are affected by the same phase noise and
accumulate a different phase shift induced by the measured field. 
Estimation of the differential phase allows high resolution thanks to noise cancellation~\cite{Stockton}.
Schemes based on this concept have resulted crucial for the precision measurement of 
rotations~\cite{Durfee}, gradients~\cite{Snadden} and fundamental constants~\cite{Fixler}. 
Differential atom interferometers have been also proposed for tests of general relativity~\cite{Dimopoulos}, 
equivalence principle~\cite{Hogan}, atom neutrality~\cite{Arvanitaki},  and for detection of gravitational waves~\cite{Dimopoulos2}.     
So far, differential interferometry \think{(DI)} has been only exploited with classical resources.
Its sensitivity is thus ultimately bounded by the shot noise (SN) limit, $\Delta \theta_{\rm SN} \approx 1/\sqrt{N}$, 
where $N$ is the number of particles in input.
For the single interferometer operation, 
a significant enhancement of phase sensitivity, up to the Heisenberg limit (HL) $\Delta \theta_{\rm HL} \approx 1/N$,  
can be obtained by using particle-entangled input 
states~\cite{SorensenNATURE2001, GiovannettiPRL2006, PezzePRL2009, HyllusPRA2012}.
This prediction is under intense experimental investigation with cold~\cite{AppelPNAS2009} 
and ultracold~\cite{EsteveNATURE2008, GrossNATURE2010, LuckeNATURE2011, BarradaNATCOMM2013,CasparPRL2013,StrobelSCIENCE2014} atoms.
However, the analysis of a single interferometer has emphasized~\cite{HuelgaPRL1997,ShajiPRA2007,EscherNATPHYS2011,DemkowiczNATCOMM2012,AcinPRL2012} 
that sub-SN cannot be reached in presence of strong phase noise. 
Is it possible to exploit DI with highly entangled states to overcome the SN~\cite{EkertPRA2006} in such a noisy environment?

%%%%%%%%%%%%%%%%%%%%%%%%%%%%%%%%%%%%%%%%%%%%%%%%%%%%%%%%%%%%%%%%%%%%%%%%%%%%%%%%%%%%%%%%%%%

\begin{figure}[!t]
\begin{center}
\includegraphics[scale=0.85]{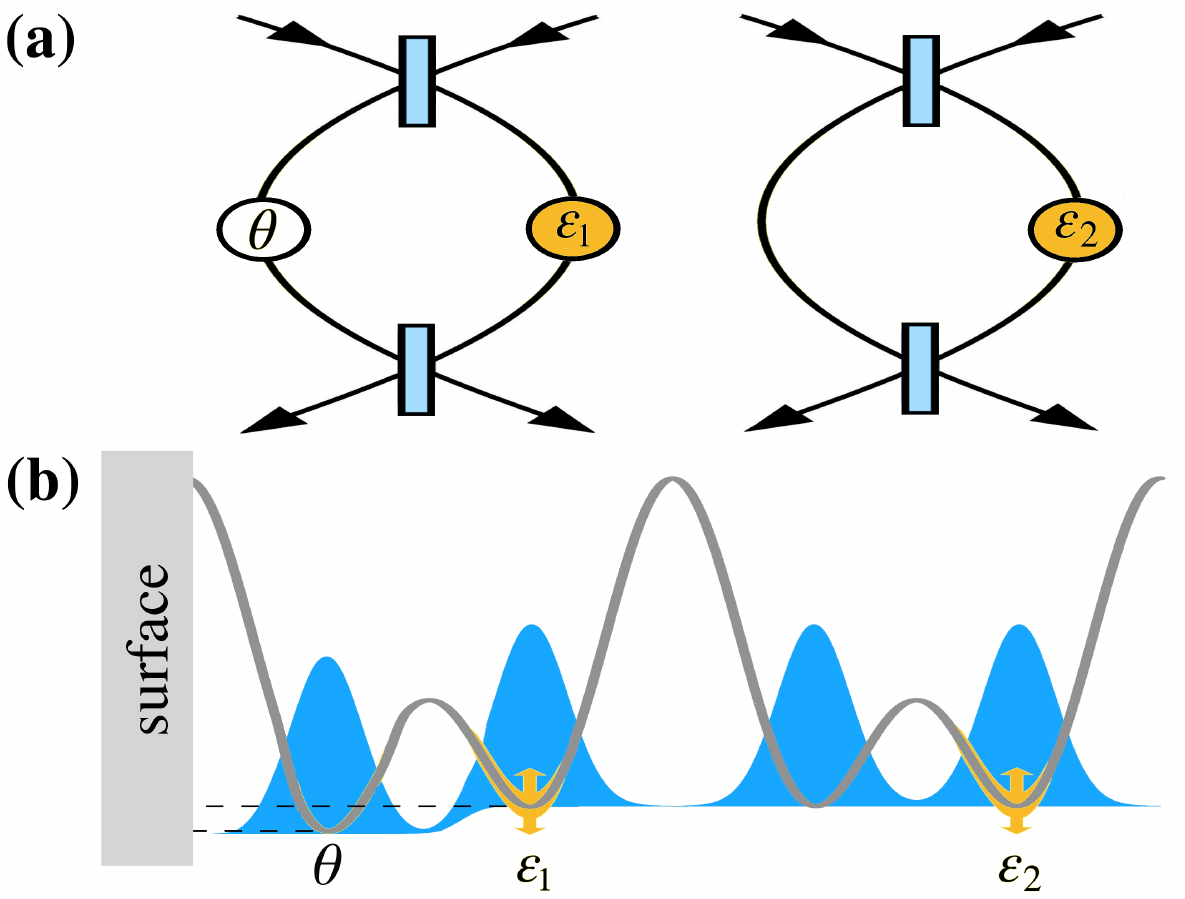}
\end{center}
\caption{\small{ 
Differential scheme discussed in this manuscript.
a) Two Mach-Zehnder interferometers
affected by shot-to-shot random phase noise $\epsilon_1$ and $\epsilon_2$. 
The signal $\theta$ can be estimated in the presence of arbitrary noise, provided that relative noise 
fluctuations are sufficiently small. 
b) Application to Bose-Einstein condensates 
(with spatial density represented by blue filled regions) trapped in a superlattice potential (grey curve). 
Splitting operations in each double-well are obtained by tuning the inter-well barrier.
Short range forces between atoms and a nearby surface induce a phase shift $\theta$. 
Trapping potential fluctuations lead to correlations between $\epsilon_1$ and $\epsilon_2$. 
}} \label{Fig:MZ} 
\end{figure} 
%%%%%%%%%%%%%%%%%%%%%%%%%%%%%%%%%%%%%%%%%%%%%%%%%%%%%%%%%%%%%%%%%%%%%%%%%%%%%%%%%%%%%%%%%%%

In this manuscript we study DI with 
two sensors implementing quantum resources
and affected by phase noise of arbitrarily large amplitude (see Fig.~\ref{Fig:MZ}) \cite{nota_optics}. 
Our analysis takes into account the correlations of the two interferometer outcomes.
It goes beyond the trivial subtraction of the two output phases estimated independently, that does not offer any significant 
quantum enhancement of phase sensitivity.
We provides the necessary and sufficient condition, based on the Fisher information, for an entangled state 
to allow sub-SN phase sensitivity. 
We also demonstrate that the HL, which is believed to be only achievable in noiseless quantum 
interferometers~\cite{HuelgaPRL1997,ShajiPRA2007,EscherNATPHYS2011,DemkowiczNATCOMM2012}, 
is preserved by the \think{lossless} differential scheme as long as relative noise fluctuations are also at the HL.
While the HL is saturated by maximally entangled states which are extremely fragile to particle losses, the SN can be overcome 
by less entangled and more robust states, as those experimentally created via particle-particle 
interaction in Bose-Einstein condensates (BECs)~\cite{
EsteveNATURE2008, GrossNATURE2010, LuckeNATURE2011, BarradaNATCOMM2013, CasparPRL2013, StrobelSCIENCE2014}. 
These findings open the door to full exploitation of quantum resources in realistic devices, provided that a DI scheme is implemented.
 
 %%%%%%%%%%%%%%%%%%%%%%%%%%%%%%%%%%%%%%%%%%%%%%%%%%%%%%%%%%%%%%%%%%

\section{Parameter estimation in differential interferometry.}

Figure \ref{Fig:MZ}(a) shows the general DI scheme discussed in this manuscript. 
It consists of two interferometers running in parallel. 
The input state $\hat \rho$ is transformed by 
$\hat U(\theta,\epsilon_1,\epsilon_2) = e^{-i (\theta+\epsilon_1) \hat{J}_1} \otimes  e^{-i \epsilon_2 \hat{J}_2}$, 
where $\hat{J}_{1,2}$ are collective spin operators for the first and second interferometer, respectively.
The phase shift in the first (second) interferometer  is $\theta+\epsilon_1$ ($\epsilon_2$),
where $\theta$ is the ``signal phase'' to be estimated and $\epsilon_1,\epsilon_2  \in [-\pi, \pi]$ is the
phase noise accumulated during the interferometer operations.
The values of $\epsilon_1$ and $\epsilon_2$ change randomly in repeated shots, %experimental realisations, 
with probability distribution $P(\epsilon_1,\epsilon_2)$. 
Our general formalism does not assume a specific noise model and 
encompasses both Markovian and non-Markovian dephasing
[we will later discuss specific forms of $P(\epsilon_1,\epsilon_2)$ \think{and focus on the case of correlated interferometer 
where $\epsilon_1 = \pm \epsilon_2$}].
We consider a general positive-operator value
measure (POVM) $\hat{E}(\mu)$ on the output state
and use an unbiased estimator $\Theta_{\rm est}(\mu_1, ..., \mu_m)$, 
which is a function of the results obtained in $m$ repeated \think{independent} measurements \cite{Helstrom}.
The variance of the estimator fulfills $\Delta \Theta_{\rm est} \geq \Delta \theta_{\rm CR}$ \cite{Cramer}, where 
\begin{equation} \label{CR}
\Delta\theta_{\rm CR} = \frac{1}{\sqrt{m F(\theta)}},
\end{equation}
is the the Cramer-Rao (CR) bound, 
\begin{equation} \label{Fisher}
	F(\theta)=\sum_{\mu} \frac{1}{P(\mu|\theta)}
	\left(\frac{d P(\mu|\theta)}{d\theta}\right)^2,
\end{equation}
is the Fisher information (FI), 
\begin{equation} \label{Eq:CondProb}
	P(\mu|\theta) = \int_{-\pi}^{\pi} d\epsilon_1  \int_{-\pi}^{\pi}  d\epsilon_2 \, 
	P(\epsilon_1,\epsilon_2) P(\mu|\theta, \epsilon_1, \epsilon_2) 
\end{equation}
are conditional probabilities, and 
$P(\mu|\theta, \epsilon_1,\epsilon_2) = {\rm Tr}[\hat{E}(\mu) \hat U(\theta,\epsilon_1,\epsilon_2) \hat{\rho} \hat U^\dag(\theta,\epsilon_1,\epsilon_2)]$. 
\think{In particular, if the state $\hat{\rho}$ is separable in the two interferometers, $\hat \rho = \hat \rho_1 \otimes \hat \rho_2$, 
and the measurement in each interferometer are independent, $\hat{E}(\mu) = \hat{E}_1(\mu_1) \otimes \hat{E}_2(\mu_2)$
[$\mu \equiv (\mu_1, \mu_2)$], then Eq.~(\ref{Eq:CondProb}) becomes
\beq \label{Eq:CondProbSep}
P(\mu_1, \mu_2 |\theta) = \int_{-\pi}^{\pi} d\epsilon_1  \int_{-\pi}^{\pi}  d\epsilon_2 \,  P(\epsilon_1,\epsilon_2)
P(\mu_1 \vert \theta + \epsilon_1) P(\mu_2 \vert \epsilon_2).
\eeq}
Equation (\ref{CR}) takes into account the full quantum correlations of the interferometers outcomes and
provides the lowest possible phase uncertainty, 
given the conditional probability distribution $P(\mu|\theta)$. 
It can be saturated for large $m$
by the maximum-likelihood estimator \cite{Cramer}. 

%%%%%%%%%%%%%%%%%%%%%%%%%%%%%%%%%%%%%%%%%%%%%%%%%%%%%%%%%%%%%%%%%%

\section{Phase sensitivity.}
\begin{figure}[!t]
\begin{center}
\includegraphics[scale=0.6]{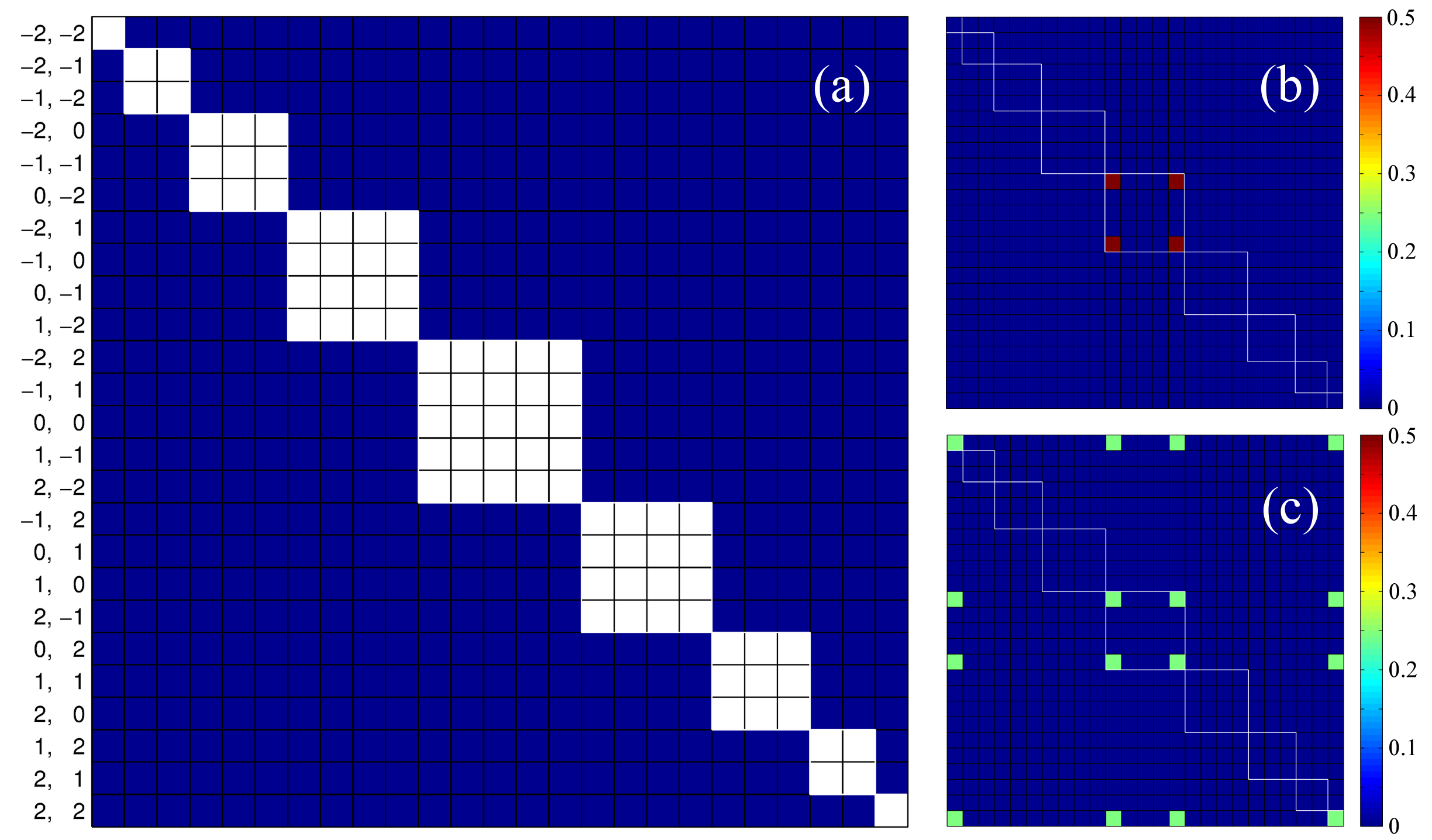}
\end{center}
\caption{\small{Panel (a) shows the general layout of a density matrix 
$\hat{\rho}$ of four particles in each interferometer. 
The basis indexes $n_1$, $n_2$ are reported on the left part of the figure. 
Following an opportune reordering of the basis, the DFS for vanishing relative noise fluctuations, 
$P_-(\epsilon_-)=\delta(\epsilon_-)$, is shown by
the white squares corresponding (from top-left to bottom-right) to $M=-4,-3,-2,..,4$.
Panel (b) shows the density matrix of state (\ref{NOON2}), fully included in the central ($M=0$) DFS.
Panel (c) shows the density matrix of state (\ref{NOON1}).
In the left panels the color scale is $\langle n_1,n_2 \vert \hat{\rho} \vert m_1, m_2 \rangle$.
}} \label{Fig:DM} 
\end{figure}

Here we calculate the highest sensitivity allowed by the above DI scheme.
We rewrite Eq. (\ref{Eq:CondProb}) as $P(\mu|\theta) = {\rm Tr}[\hat{E}(\mu) \hat U(\theta) \hat{\rho}_{\rm eff} \hat U^\dag(\theta)]$,
where $\hat U(\theta) \equiv e^{-i \theta \hat J_1} \otimes {\ensuremath{\mathbbm 1}}_2$ depends solely on $\theta$ and 
\be
\hat{\rho}_{\rm eff} \equiv \int_{-\pi}^{\pi} d\epsilon_1 d\epsilon_2 \, P(\epsilon_1,\epsilon_2) \, 
\hat U(0,\epsilon_1,\epsilon_2) \, \hat{\rho} \, \hat U^\dag(0,\epsilon_1,\epsilon_2).
\ee
The noisy differential interferometer with input $\hat{\rho}$ is thus equivalent to a noiseless interferometer 
with effective input density matrix $\hat{\rho}_{\rm eff}$.
This equivalence can be used to \think{minimize}
$\Delta \theta_{\rm CR}$ over all possible POVMs~\cite{BraunsteinPRL1994, Helstrom}.
We have 
\be
\Delta\theta_{\rm CR} \geq \frac{1}{\sqrt{m F_Q[\hat{\rho}_{\rm eff}]}},
\ee
where $F_Q[\hat{\rho}_{\rm eff}]=4 (\Delta \hat{R})^2$ is the quantum Fisher information (QFI) and $\hat{R}$ is obtained by solving 
$\{\hat{R}, \hat{\rho}_{\rm eff}\}=i[\hat{\rho}_{\rm eff}, \hat{J}_1 \otimes {\ensuremath{\mathbbm 1}}_2]$.
Taking ${\vert n_{i} \rangle}$ the eigenbasis of $\hat{J}_{i}$
[$\hat{J}_{i} \vert n_{i} \rangle = n_{i} \vert n_{i} \rangle$, $-N_{i}/2 \leq n_{i} \leq N_{i}/2$], where $i=1,2$ labels the interferometer,
we have
\be
\langle n_1,n_2 \vert \hat{\rho}_{\rm eff} \vert m_1,m_2 \rangle = C^{n_1,n_2}_{m_1,m_2} \langle n_1,n_2 \vert \hat{\rho} \vert m_1,m_2 \rangle,
\ee 
where 
\be \label{C}
C^{n_1,n_2}_{m_1,m_2} = \int_{-\pi}^{\pi} d\epsilon_1 d\epsilon_2  P(\epsilon_1,\epsilon_2) 
e^{-i [\epsilon_1(n_1-m_1) + \epsilon_2(n_2-m_2)]}. 
\ee
Depending on $P(\epsilon_1,\epsilon_2)$, the DI may admit a decoherence free subspace (DFS)
spanned by states of the system that experience no evolution under the noise~\cite{noteDFS}.
For uncorrelated noise [$P(\epsilon_1,\epsilon_2)=P_1(\epsilon_1)P_2(\epsilon_2)$]
we obtain $C^{n_1,n_2}_{m_1,m_2}=1$ if and only if $n_{1,2}=m_{1,2}$, i.e. 
the DFS simply reduces to the eigenstates of $\hat{J}_{1,2}$. 
These states are insensitive to the phase shift and thus useless for phase estimation.
As common in several differential atom interferometers,
we assume that $P(\epsilon_1,\epsilon_2)=P_+(\epsilon_+)P_-(\epsilon_-)$, where $\epsilon_{\pm}=(\epsilon_1 \pm \epsilon_2)/2$
indicates the total (``+'' sign) and relative (``-'' sign) noise.
Equation~(\ref{C}) becomes 
\be
C^{n_1,n_2}_{m_1,m_2} = \tilde{P}_+(n_1-m_1+n_2-m_2) \tilde{P}_-(n_1-m_1-n_2+m_2), \nonumber
\ee
where $\tilde{P}_\pm(k) \equiv \int_{-\pi}^{\pi} d\epsilon P_\pm(\epsilon) e^{-i k \epsilon}$.
A non-trivial DFS, defined by the condition $n_1+n_2=m_1+m_2$ [$n_1-n_2=m_1-m_2$], 
exists \think{in the limit of} vanishing relative $P_-(\epsilon_-)=\delta(\epsilon_-)$ [total, $P_+(\epsilon_+)=\delta(\epsilon_+)$] noise fluctuations. 
Such DFS can be decomposed in subspaces defined by constant values of $M=n_1+n_2$ [$M=n_1-n_2$].
These, except the trivial case $M=\pm (N_1+N_2)/2$, contain coherence terms and are thus relevant for phase estimation.
For vanishing total (relative) noise and in the presence of large relative (total) noise, $\hat \rho_{\rm eff}$ becomes block diagonal,
\think{[see Fig.~\ref{Fig:DM},a] describing a statistical mixture of states with definite $M$ values.
We write} $\hat \rho_{\rm eff} = \sum_M Q_M \hat{\rho}_M$, where 
$\hat \rho_M = \hat{\pi}_M \hat{\rho} \hat{\pi}_M$, 
$\hat{\pi}_M$ are projectors into the fixed-$M$ subspace and 
$Q_M = {\rm Tr}[\hat{\pi}_M \hat{\rho} \hat{\pi}_M]$ are weights satisfying $\sum_M Q_M=1$.
We have $F_Q[\hat \rho_{\rm eff}] \leq \sum_M Q_M F_Q[\hat \rho_M] \leq 4\sum_M Q_M (\Delta \hat{J}_1)^2_M$, 
where $(\Delta \hat{J}_1)^2_M$ is the variance of $\hat{J}_1$ calculated for $\hat \rho_M$ and the second 
bound can be saturated by pure states.
For separable states, following \cite{PezzePRL2009} and assuming, for simplicity, $N_1=N_2=N$, we have
\begin{equation} \label{sep}
	F_Q[\hat \rho_{\rm eff}] \leq N - \sum_{M=\tilde{M}}^{N} (Q_M+Q_{-M}) [N-(N-M)^2] \leq N,
\end{equation}
where $\tilde{M}$ is the solution of $(N-M)^2=N$.
In general, 
\begin{equation} \label{ent}
	F_Q[\hat \rho_{\rm eff}] \leq N^2 - \sum_{M=1}^N (Q_M+Q_{-M}) (2N - M) M \leq N^2.
\end{equation}
In Eqs.~(\ref{sep}) and (\ref{ent}), the QFI is maximized by populating only the $M=0$ subspace.
We recover the same phase uncertainty bounds as in 
ideal noiseless case:  the SN limit, 
$\Delta\theta_{\rm SN} = 1/\sqrt{m N}$, for separable states, and the HL,
$\Delta\theta_{\rm HL} = 1/\sqrt{m}N$, for general quantum states.
In other words, the condition 
$F_Q[\hat{\rho}_{\rm eff}] > N$,
is necessary and sufficient for reaching sub-SN sensitivities.  
Moreover,  according to Eq.~(\ref{sep}),
overcoming the SN necessarily requires particle entanglement in the effective input state.
In full analogy to the noiseless case, there exist optimal entangled states providing a quadratic enhancement of phase sensitivity
even in presence of large phase noise.
The optimal states for DI are
[in the $\vert m_1,m_2 \rangle$ basis, see Fig.~\ref{Fig:DM},b]
\begin{equation}  \label{NOON2}
\label{cases}
\vert \psi_{\rm opt} \rangle = \cases{\frac{|N/2,-N/2\rangle+|-N/2,N/2\rangle}{\sqrt{2}} &if $P_-(\epsilon_-)=\delta(\epsilon_-)$\\
\frac{|N/2,N/2\rangle+|-N/2,-N/2\rangle}{\sqrt{2}} &if $P_+(\epsilon_+)=\delta(\epsilon_+)$.\\}
\end{equation}
These have $F_Q = N^2$ and are not affected by phase noise. 
The states (\ref{NOON2}) have been experimentally realized with 2 \cite{RoosNATURE2006} 
and up to 8 \cite{MonzPRL2011} trapped ions, and further investigated in \cite{DornerNJP2012}.
While the saturation of the HL requires entangled interferometers, 
we can still have a HL scaling, i.e. $F_Q \propto N^2$ 
if we consider states which are separable in the two interferometers, $\hat \rho = \hat \rho_1 \otimes \hat \rho_2$
A prominent example is the product of NOON states, 
\be \label{NOON1}
\vert \psi_{\rm NOON} \rangle = \bigg( \frac{|N/2\rangle+|-N/2\rangle}{\sqrt{2}} \bigg)_1 \otimes \bigg( \frac{|N/2\rangle+|-N/2\rangle}{\sqrt{2}} \bigg)_2,
\ee 
which, as shown in Fig.~\ref{Fig:DM},c, does not \think{(entirely)} belong to the DFS and has $F_Q = N^2/4$ 
both when $P_-(\epsilon_-)=\delta(\epsilon_-)$ and when $P_+(\epsilon_+)=\delta(\epsilon_+)$.

%%%%%%%%%%%%%%%%%%%%%%%%%%%%%%%%%%%%%%%%%%%%%%%%%%%%%%%%%%%%%%%%%%%

%%%%%%%%%%%%%%%%%%%%%%%%%%%%%%%%%%%%%%%%%%%%%%%%%%%%%%%%%%%%%

\section{Differential interferometry with NOON states} \label{Sec.NOON}

In this section we study the differential interferometer scheme 
$\hat{U}(\theta,\epsilon_1,\epsilon_2)= (e^{-i \frac{\pi}{2} \hat J_x} e^{-i (\theta+\epsilon_1) \hat J_z})_1 \otimes (e^{-i \frac{\pi}{2} \hat J_x} e^{-i \epsilon_2 \hat J_z})_2$, 
each interferometer being represented by the unitary transformation given by a 
phase shift rotation around the $z$ axis followed by a 50-50 beam splitter. 
We further assume the phase noise distribution $P(\epsilon_1,\epsilon_2)=P_+(\epsilon_+) P_-(\epsilon_-)$.
As input state we take the direct product of NOON states, 
$\vert \psi_{\rm NOON} \rangle_z \equiv \vert \rm{NOON} \rangle_z \otimes \vert \rm{NOON} \rangle_z$,
with components along the $z$ direction, $\vert {\rm NOON} \rangle_z = (\vert N/2 \rangle_z + \vert -N/2 \rangle_z)/\sqrt{2}$ \think{\cite{BollingerPRA1996}}, 
$\vert \mu \rangle_z$ being an eigenstate of $\hat{J}_z$ with eigenvalue $\mu$.
\think{In the following we provide the conditional probabilities and FI
when $P_\pm(\epsilon_{\pm})$ are even functions of $\epsilon_\pm$ and specialize to the case a Gaussian noise distributions.  
For a discussion on more general noise functions see Appendix C.}

%%%%%%%%%%%%%%%%%%%%%%%%%%%%%%%%%%%%%%%%%%%%%%%%%%%%%%%%%%%%%%%%%%%%%%%%%%%%%%%%%%%%%%%%%%%
\begin{figure}[!t]
\begin{center}
\includegraphics[scale=0.57]{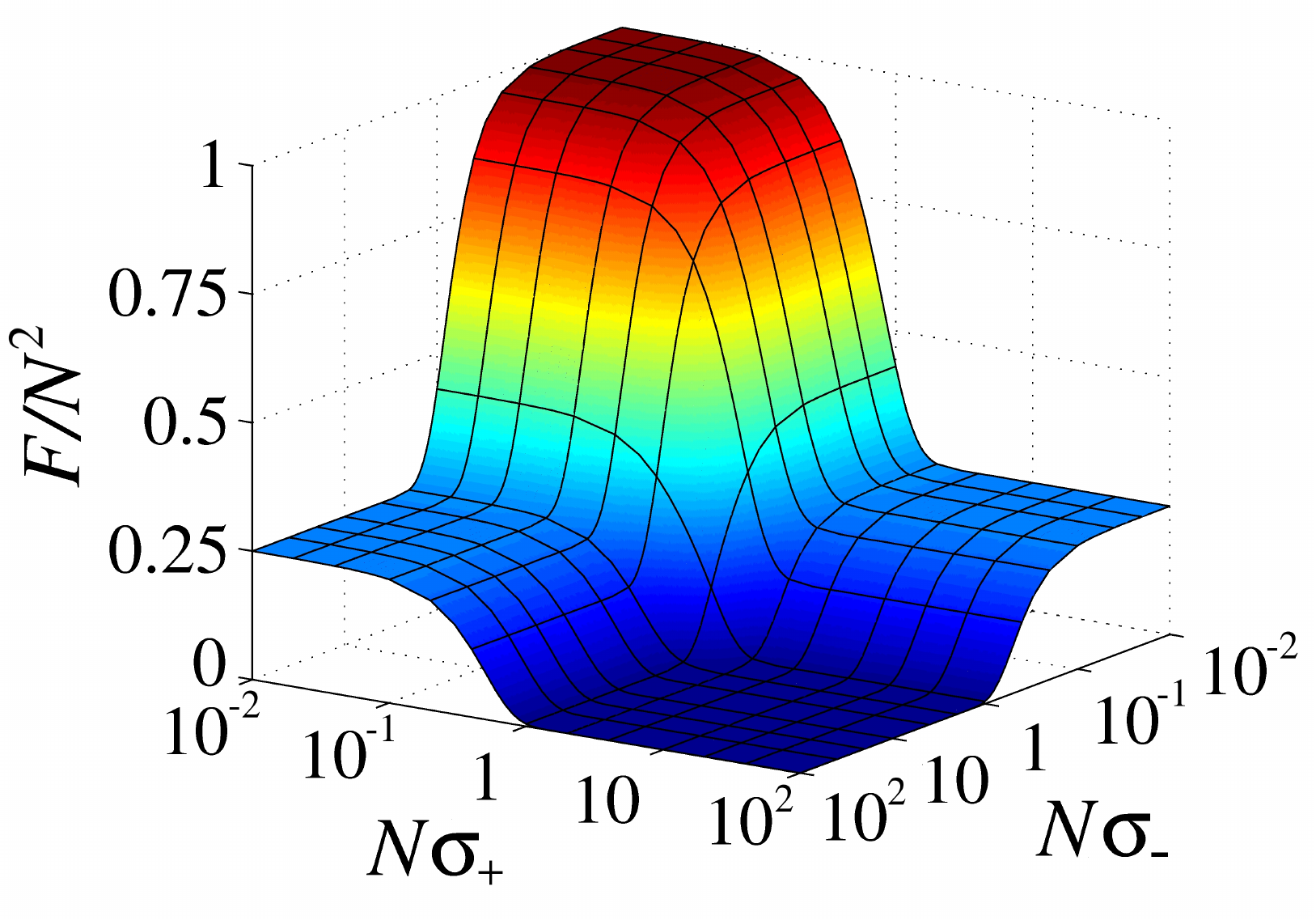}
\end{center}
\caption{\small{  
\think{Fisher information (maximized over $\theta$) for a differential interferometer with noise distribution 
given by Eq.~(\ref{NoiseDistP}) and factorized NOON states Eq.~(\ref{NOON1}) as input} (see text for details).
Here $N=100$.
}} \label{Fig:NOON} 
\end{figure} 
%%%%%%%%%%%%%%%%%%%%%%%%%%%%%%%%%%%%%%%%%%%%%%%%%%%%%%%%%%%%%%%%%%%%%%%%%%%%%%%%%%%%%%%%%%%
\think{The probability to measure a relative number of particles $\mu_1$ at the output of the interferometer 1 and 
$\mu_2$ at the output of interferometer 2} can be calculated analytically and is given by 
\think{(see Appendix C for details on the derivation of the equations below)}
\beq \label{Probability.NOON}
&& P(\mu_1, \mu_2 \vert \theta) = \bigg( \frac{N!}{2^N} \bigg)^2 
\frac{ 1 } { (\frac{N}{2}-\mu_1)! (\frac{N}{2}+\mu_1)! (\frac{N}{2}-\mu_2)! (\frac{N}{2}+\mu_2)!} \times \nonumber \\
&& \quad \,\, \times \Bigg[ 1 + (-1)^{\mu_2} V_N^+ V_N^-  + (-1)^{\mu_1} \bigg( V_N^+ V_N^- + 
\frac{(-1)^{\mu_2}}{2} \big(V_{2N}^+ + V_{2N}^-\big) \bigg) \cos N\theta  \Bigg], \nonumber \\
\eeq
where $V^{\pm}_K \equiv \int_{-\pi}^{\pi} d \epsilon_\pm P_\pm(\epsilon_\pm) \cos(K \epsilon_\pm)$.
The FI is 
\be \label{Fisher.NOON}
F(\theta) = \frac{N^2 \sin^2 N\theta}{2} \bigg[ \frac{A_- B_-^2}{A_-^2 - B_-^2 \cos^2 N\theta} + \frac{A_+ B_+^2}{A_+^2 - B_+^2 \cos^2 N\theta} \bigg],
\ee
where $A_\pm(N) \equiv 1 \pm V_N^+ V_N^-$, $B_\pm(N) \equiv V_N^+ V_N^- \pm (V_{2N}^+ + V_{2N}^-)/2$.
The optimal value of the FI\think{, $F \equiv \max_{\theta} F(\theta)$,} is reached for $\cos N\theta = 0$,
\be \label{FIopt}
F = \frac{N^2}{2} \left[ \frac{B_-^2(N)}{A_-(N)} + \frac{B_+^2(N)}{A_+(N)} \right].
\ee 
Let us discuss the different limit values of Eq.~(\ref{FIopt}) taking into account that 
$V^{\pm}_K=1$ when $P_\pm(\epsilon_\pm) = \delta(\epsilon_\pm)$
and $V^{\pm}_K=0$ when $P_\pm(\epsilon_\pm) = 1/2\pi$.
If relative noise fluctuations are vanishingly small, $P_- (\epsilon_-) = \delta(\epsilon_-)$, 
Eq.~(\ref{FIopt}) ranges from $F=N^2$ 
[if also $P_+ (\epsilon_+) = \delta(\epsilon_+)$, corresponding to the ideal noiseless limit]
to $F=N^2/4$ [when $P_+ (\epsilon_+) = 1/2\pi$].
In other words, if the relative noise between the two interferometers is fixed, 
a phase sensitivity at the HL can be obtained for arbitrary large total noise 
(i.e. arbitrary large noise in {\it each} interferometer).
If total noise fluctuations are large, $P_+ (\epsilon_+) = 1/2\pi$, 
we obtain $F=N^2 (V_{2N}^-)^2$, which predicts the HL for $V_{2N}^- \approx 1$
and sub-SN for $V_{2N}^- > 1/\sqrt{N}$.
In Fig.~\ref{Fig:NOON}(c) we plot Eq.~(\ref{FIopt}) as a function of $\sigma_{\pm}$
taking 
\beq \label{NoiseDistP}
P(\epsilon_\pm) = \frac{e^{(\cos \epsilon_\pm)/\sigma_\pm^2}}{2 \pi I_0(1/\sigma_\pm^2)},
\eeq
where $I_0(x)$ is the modified Bessel function of the first kind.
\think{This noise function continuously interpolates from a Gaussian distribution of width $\sigma_\pm$, when $\sigma_\pm \ll 1$, 
to a flat distribution, when $\sigma_\pm \gg 1$.}
The condition $V_{2N}^- \approx 1$ is thus equivalent to $\sigma_- \approx 1/N$, while 
$V_{2N}^- \aplt 1/\sqrt{N}$ is recovered for $\sigma_- \apgt \sqrt{\log N}/N$. 
These results show that  reaching the HL in the differential interferometer requires relative noise fluctuations at the HL itself.

%%%%%%%%%%%%%%%%%%%%%%%%%%%%%%%%%%%%%%%%%%%%%%%%%%%%%%%%%%%%%%%%%%%%%%%%%%%%%%%%%%%%%%%%%%%

\section{Precision limit for DI with Bose-Einstein condensates.} \label{Sec:BEC}
In this section we discuss a differential Mach-Zehnder (MZ) interferometer 
$\hat U(\theta, \varepsilon_1, \varepsilon_2) = (e^{-i (\theta + \varepsilon_1) \hat{J}_y})_1 \otimes (e^{-i \varepsilon_2 \hat{J}_y})_2$ with
input states that can be created with two-mode BECs. 
We consider phase estimation from the measurement of the number of particles in output. 
We further take \think{vanishing relative phase noise fluctuations in the two interferometers,} $P_- (\epsilon_-) = \delta(\epsilon_-)$, and 
$P(\epsilon_+)$ given by Eq.~(\ref{NoiseDistP}).
%$P(\epsilon_+) = \frac{e^{(\cos \epsilon_+)/\sigma_+^2 }}{2 \pi I_0(1/\sigma_+^2)}$.
 
%%%%%%%%%%%%%%%%%%%%%%%%%%%%%%%%%%%%%%%%%%%%%%%%%%%%%%%%%%%%%%%%%%%%%%%%%%%%%%%%%%%%%%%%
\begin{figure}[!b]
\begin{center}
\includegraphics[scale=0.42]{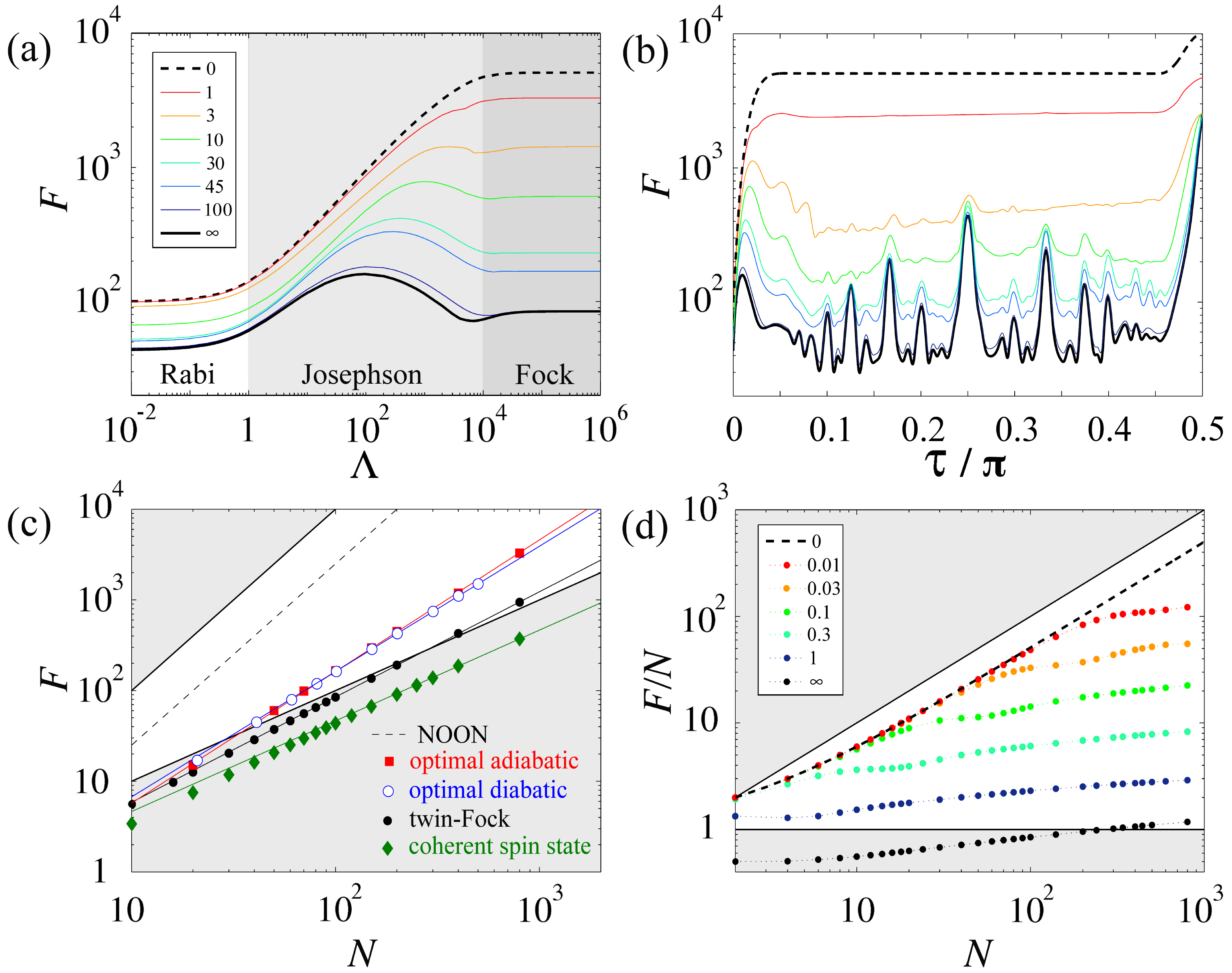}
\end{center}
\caption{\small{ 
a) FI as a function of $\Lambda$ for the adiabatic state preparation.
The shadow regions highlight different regimes (see text). 
b) FI as a function of $\tau$ for the diabatic state preparation.
In panels (a) and (b) $N=100$ and different lines refer to different values of $N \times \sigma_+$.
c) FI as a function of $N$ for various input states of the 
differential MZ interferometer for $\sigma_+=\infty$.
Solid thin lines are fits while solid thick lines are the HL and SN (the white region
between the two lines corresponds to sub-SN). 
d) FI for the twin-Fock state as a function of $N$ and for different $\sigma_+$ values.
Think lines and colored regions are as in panel (c).
}} \label{Fig:adiabatic} 
\end{figure} 
%%%%%%%%%%%%%%%%%%%%%%%%%%%%%%%%%%%%%%%%%%%%%%%%%%%%%%%%%%%%%%%%%%%%%%%%%%%%%%%%%%%%%%%%

We first consider adiabatic state preparation~\cite{JavanainenPRA1999}, focusing on the ground state
$\vert \psi_{\rm gs}(\Lambda)\rangle$ of the Hamiltonian 
$\hat{H} = \hbar \chi \hat{J}_z^2 - \hbar \Omega \hat{J}_x$, with $\Lambda \equiv N\chi/\Omega$.
This can be implemented in a double-well trap [see Fig. \ref{Fig:MZ}(b)] with %, in which case
$\chi$ and $\Omega$ %are the (positive) 
interaction and tunneling parameters, respectively~\cite{EsteveNATURE2008, BarradaNATCOMM2013}.
We can distinguish Rabi, $0 < \Lambda \leq 1$,
Josephson, $1 < \Lambda \leq N^2$, and Fock, $\Lambda > N^2$, regimes. 
In the ideal case, these regimes are characterized by different scalings of the 
FI (optimized in $\theta$): $F \sim N$ ($F=N$ for the spin coherent state, $\vert N/2 \rangle_x$, $\Lambda=0$), 
$F \sim N^{3/4}$, and $F\sim N^2$ [$F=N^2/2+N$ \cite{Pezze2013} for the twin-Fock state, $\vert 0\rangle_z$, $\Lambda=\infty$], 
respectively~\cite{PezzePRA2005}. 
In Fig.~\ref{Fig:adiabatic}(a) we report the FI as a function of $\Lambda$ for 
the differential MZ with input $\vert \psi_{\rm gs}(\Lambda)\rangle \otimes \vert \psi_{\rm gs}(\Lambda)\rangle$.
Different lines refer to different value of $N \times \sigma_+$, ranging from the noiseless case 
($\sigma_+=0$, thick dashed line) to uniform phase noise ($\sigma_+=\infty$, thick solid line).
The twin-Fock is optimal for $\sigma_+ \apgt 1/N$, while for $\sigma_+ \to \infty$
 the FI is maximized at $\Lambda\approx N$. 
  
We further consider states that are created by the nonlinear 
evolution $\vert \psi_{\rm dyn}(\tau)\rangle = e^{i \delta \hat{J}_n} \, e^{-i\hat{J}_z^2 \tau}  \vert N/2 \rangle_x$ 
starting from a spin coherent state \cite{GrossNATURE2010, SorensenNATURE2001, KitagawaPRA1993}, where $\tau=\chi t$.
Here $e^{i \delta \hat{J}_n}$ rotates the state so to maximize the FI in the noiseless case \cite{PezzePRL2009,notaFerrini}.
In Fig~\ref{Fig:adiabatic}(b) we show the FI for
the differential MZ with input $\vert \psi_{\rm dyn}(\tau)\rangle \otimes \vert \psi_{\rm dyn}(\tau)\rangle$.
Lines are as in Fig.~\ref{Fig:adiabatic}(a).
Interestingly, the short time dynamics ($\tau \apgt 1/\sqrt{N}$, where spin squeezing is created \cite{KitagawaPRA1993}) 
is robust and sub-SN is found also for large phase noise.
The characteristic plateau ($F=N^2/2$ for the ideal case \cite{PezzePRL2009}) 
is washed out when $\sigma_+ \aplt 1/N$.
The FI at large values of $\sigma_+$ is characterized by several peaks, the most prominent found in correspondence to the creation of
macroscopic superposition states (``phase cats'') with multiple (larger than two) components.
The long-time dynamics at $\tau=\pi/2$ leads to maximally entangled states (a two components phase cat)
having $F=N^2/4$, as discussed above.

For specific input states, we have repeated the previous analysis for large values of $N$ (up to $N \approx 1000$) and 
$\sigma_+=\infty$ (flat total noise case). 
The results are shown in Fig.~\ref{Fig:adiabatic}{(c)}.
The FI reaches an asymptotic power law scaling $F=\beta N^\alpha$:
$\beta=0.5$, $\alpha = 1$ for coherent spin state (green diamonds);
$\beta=0.2$, $\alpha=1.5$ for the optimal states of the adiabatic preparation at $\Lambda\approx N$ (red squares);
$\beta=0.3$, $\alpha=1.4$ for the optimal states of the diabatic preparation at time $\tau \approx 1/N^{3/4}$ (blue circles);
$\beta=0.39, \alpha=1.17$ for the twin-Fock state {(black dots)}. 
The solid black line is the analytical NOON state result ($\alpha=2$, $\beta=1/4$) discussed previously.
The twin-Fock state is an interesting and experimentally relevant \cite{LuckeNATURE2011} example.
It is strongly entangled and reaches a HL scaling 
in the single noiseless MZ, however it performs only slightly better than the SN in the differential MZ
with large noise and large number of particles.
In Fig~\ref{Fig:adiabatic}(d) we further investigated the FI for the twin-Fock state 
as a function of $N$, for different values of $\sigma_+$ (dots).
For $N \apgt 1/\sigma_+$ and $\sigma_+\ll1$, the FI follows the ideal behavior $F=N^2/2+N$ (dashed line). 
For $N \gg 1/\sigma_+$, we recover roughly the same scaling of FI ($F \propto N^{1.17}$) as in the large phase noise case.%, yet with a 
%prefactor increasing with .

%%%%%%%%%%%%%%%%%%%%%%%%%%%%%%%%%%%%%%%%%%%%%%%%%%%%%%%%%%%%%%%%%%

\section{Conclusions.}
In this manuscript we have extended the analysis of DI to the domain of entangled states. 
It is not obvious, a priori, that DI can suppresses spurious phase noise when 
highly entangled -- and thus extremely fragile against \think{phase} noise fluctuations -- states are used.
\think{Our analysis reveals that when the phase noise is perfectly correlated in the two interferometers, and losses can be neglected, 
there exists a decoherence free subspace where entanglement is passively protected.}  
We have thus identified a class of entangled input state that can provide a sub-SN sensitivity in
a differential interferometer up to the HL, even for large noise. 
This class is non trivial, fully characterized by the FI, 
and includes states that have been recently created experimentally.
We expect our results to be a guideline for 
quantum-enhanced realistic interferometers in the near future.

\section{Acknowledgment.}
This work has been supported by the EU-STREP Project QIBEC and 
ERC StG No. 258325. L.P. and M.F. acknowledge financial support by MIUR through FIRB Project No. RBFR08H058.

%%%%%%%%%%%%%%%%%%%%%%%%%%%%%%%%%%%%%%%%%%%%%%%%%%%%%%%%%%%%%%%%%%

\appendix

\section{Derivation of Eq.~(\ref{Eq:CondProb}).} 
\label{Sec:1}

We here derive Eq.~(\ref{Eq:CondProb}) from first principles.
We start from the joint probability density $P(\mu, \theta, \epsilon_1, \epsilon_2)$ and integrate over $\epsilon_1$ and $\epsilon_2$,
\be \label{Prob}
P(\mu, \theta) = \int_{-\pi}^{\pi} d \epsilon_1 d \epsilon_2 \, P(\mu, \theta, \epsilon_1, \epsilon_2) 
\ee
so to eliminate nuisance parameters.
By using the relation $P(x,y)=P(x\vert y) P(y)$ between joint and conditional probabilities, where $x$ and $y$ are random variables,
we have 
\be
P(\mu \vert \theta)  = \frac{P(\mu, \theta)}{P(\theta)} = \int_{-\pi}^{\pi} d \epsilon_1 d \epsilon_2 \, P(\mu \vert \theta, \epsilon_1, \epsilon_2)
P(\epsilon_1, \epsilon_2 \vert \theta).
\ee
Since $\epsilon_{1,2}$ do not depend on $\theta$, i.e. $P(\epsilon_1,\epsilon_2 \vert \theta)=P(\epsilon_1, \epsilon_2)$, we recover Eq. (\ref{Eq:CondProb}).

%%%%%%%%%%%%%%%%%%%%%%%%%%%%%%%%%%%%%%%%%%%%%%%%%%%%%%%%%%%%%%%%%%

\section{Derivation of inequalities~(\ref{sep}) and (\ref{ent}).}

Here we detail the calculation of $4\sum_M Q_M (\Delta \hat{J}_1)^2_M$, where $(\Delta \hat{J}_1)^2_M$ is the variance of 
the operator $\hat{J}_1$ calculated on the fixed-$M$ subspace.
We consider the case $N_1=N_2 =N/2$ for simplicity. 
We have $4 (\Delta \hat{J}_1)^2_M \leq \big( \max_M n_1 -  \min_M n_1 \big)^2$,
where $-N/2 \leq n_1 \leq N/2$ are the eigenvalues of $\hat{J}_1$ and 
$\max_M n_1$ ($\min_M n_1$) are the maximum (minimum) values of $n_1$ in the fixed-$M$ DFS.
In general, $\max_M n_1 = \min(M+N/2,N/2)$ and $\min_M n_1 = \max(M-N/2,-N/2)$.
We thus have
\be
4 (\Delta \hat{J}_1)^2_M \leq \big( N-\vert M \vert \big)^2,
\ee
with equality of $\rho_M$ is the equal-weighted superposition of states with the maximum and the minimum value of $n_1$
in the fixed-$M$ DFS.
Taking into account that $\sum_{M=-N}^N Q_M = 1$
\beq
4\sum_M Q_M (\Delta \hat{J}_1)^2_M &=& Q_0 N^2 + \sum_{M=1}^N (Q_M+Q_{-M}) (N-M)^2 \nonumber \\
&=& N^2 - \sum_{M=1}^N (Q_M+Q_{-M}) M (2N-M) \nonumber
\eeq
We thus recover Eq.~(\ref{ent}).
Since $M \leq 2N$, the second term in the equation above is nonnegative and we find $4\sum_M Q_M (\Delta \hat{J}_1)^2_M \leq N^2$. For separable states, we need to further take into account that 
$4 (\Delta \hat{J}_1)^2_M \leq N$ \cite{PezzePRL2009}. We thus have 
\be
4 (\Delta \hat{J}_1)^2_M \leq \min\big[ \big( N-\vert M \vert \big)^2, N\big],
\ee
and thus obtain
\beq
4\sum_M Q_M (\Delta \hat{J}_1)^2_M &=& Q_0 N +  \sum_{M=1}^N (Q_M+Q_{-M}) \min\big[ (N-M)^2, N\big] \nonumber \\
&=& Q_0 N - \sum_{M=\tilde{M}}^{N} (Q_M+Q_{-M}) \big[ N- (N-M)^2 \big], \nonumber 
\eeq 
which follows since for $\tilde{M} \leq M \leq N$ we have $\min[ N, (N-M)^2 ] = (N-M)^2$. We recover Eq.~(\ref{sep}).
Since for $\tilde{M} \leq M \leq N$ we have $N- (N-M)^2 \geq 0$, the second term in the above equation is
nonnegative and we have $4\sum_M Q_M (\Delta \hat{J}_1)^2_M \leq N$ for separable states.

%%%%%%%%%%%%%%%%%%%%%%%%%%%%%%%%%%%%%%%%%%%%%%%%%%%%%%%%%%%%%

\think{\section{Extension of the discussion of Sec.~\ref{Sec.NOON} to arbitrary noise distributions.}

Here we provide a detailed derivation of the equations presented in Sec.~\ref{Sec.NOON}
and extend the discussion to arbitrary noise distributions.
Let us first calculate the conditional probability distribution of the relative number of particles for the single interferometer
with a NOON probe state:
\beq
P(\mu \vert \phi) &=& \big\vert {}_z\langle \mu \vert e^{-i \frac{\pi}{2} \hat J_x} e^{-i \phi \hat J_z} \vert {\rm NOON} \rangle_z \big\vert^2 \nonumber \\
&=& \bigg\vert \frac{  {}_z\langle \mu \vert e^{-i \frac{\pi}{2} \hat J_x} \vert + N/2 \rangle_z  +  
e^{i \phi N} {}_z\langle \mu \vert e^{-i \frac{\pi}{2} \hat J_x} \vert -N/2 \rangle_z 
 }{\sqrt{2}} \bigg\vert^2. \nonumber
\eeq
The rotation matrix elements ${}_z\langle \mu \vert e^{-i \frac{\pi}{2} \hat J_x} \vert \pm N/2 \rangle_z$ 
are given by
\beq
{}_z\langle \mu \vert e^{-i \frac{\pi}{2} \hat J_x} \vert + N/2 \rangle_z &=& 
\frac{e^{-i \frac{\pi}{2}(\mu -\frac{N}{2})}}{2^{N/2}} \sqrt{\frac{N!}{(N/2-\mu)! (N/2+\mu)!}}, \nonumber \\
{}_z\langle \mu \vert e^{-i \frac{\pi}{2} \hat J_x} \vert -N/2 \rangle_z &=& 
\frac{e^{-i \frac{\pi}{2}(\mu +\frac{N}{2})}}{2^{N/2}} \sqrt{\frac{N!}{(N/2+\mu)! (N/2+\mu)!}} (-1)^{\frac{N}{2}+\mu}. \nonumber
\eeq
We thus obtain 
\beq \label{CondProbNOON}
P(\mu \vert \theta) = \frac{1}{2^N} \frac{N!}{(N/2-\mu)! (N/2+\mu)!} \,
\Bigg\vert \frac{e^{-i \frac{N}{2}\phi} + (-1)^{\mu} e^{i \frac{N}{2}\phi}}{\sqrt{2}} \Bigg\vert^2, 
\eeq
with 
\beq
\Bigg\vert \frac{e^{-i \frac{N}{2}\phi} + (-1)^{\mu} e^{i \frac{N}{2}\theta}}{\sqrt{2}} \Bigg\vert^2 =
\left\{ 
  \begin{array}{l l}
    1+(-1)^\mu \cos N \phi & \quad {\rm if}\,N\,{\rm is\,even},\\
    1+(-1)^{\mu+1/2} \sin N \phi & \quad {\rm if}\,N\,{\rm is\,odd}.
  \end{array} \right. \nonumber
\eeq
We now consider the differential sensor described by the unitary operator  
$\hat{U}(\theta,\epsilon_1,\epsilon_2)= (e^{-i \frac{\pi}{2} \hat J_x} e^{-i (\theta+\epsilon_1) \hat J_z})_1 \otimes 
(e^{-i \frac{\pi}{2} \hat J_x} e^{-i \epsilon_2 \hat J_z})_2$, 
each interferometer being given by the transformation 
$(e^{-i \frac{\pi}{2} \hat J_x} e^{-i (\phi_i) \hat J_z})_i$, $i=1,2$.
We take a NOON state of $N$ particles as input of each interferometer 
(without loss of generality we assume $N$ to be even)
and estimate the phase shift from the measurement of the relative number of particles at the output ports of each interferometer, 
$\hat{E}(\mu) \equiv \hat{E}(\mu_1,\mu_2) = 
\big( \vert \mu_1 \rangle_z \langle \mu_1 \vert \big)_1 \otimes \big( \vert \mu_2 \rangle_z \langle \mu_2 \vert \big)_2$.
Taking $P(\epsilon_1,\epsilon_2)=P_+(\epsilon_+) P_-(\epsilon_-)$, where 
$\epsilon_{\pm}=(\epsilon_1 \pm \epsilon_2)/2$, Eq.~(\ref{Eq:CondProbSep}) writes
\beq
P(\mu_1, \mu_2 \vert \theta) = \int_{-\pi}^{\pi} d \epsilon_+ P_+(\epsilon_+)
\int_{-\pi}^{\pi} d \epsilon_- P_-(\epsilon_-)
P(\mu_1 \vert \theta+\epsilon_+ +\epsilon_-) P(\mu_2 \vert \epsilon_+ - \epsilon_-), \nonumber
\eeq
with $P(\mu_i \vert \phi_i)$ ($i=1,2$) given by Eq.~(\ref{CondProbNOON}).
After straightforward algebra we obtain
\beq
P(\mu_1, \mu_2 \vert \theta) =
\bigg( \frac{N!}{2^N} \bigg)^2 
\frac{ \mathcal{A}_N(\mu_2) + \mathcal{C}_N(\mu_1, \mu_2) \cos N \theta - \mathcal{S}_N(\mu_1, \mu_2) \sin N \theta} { (\frac{N}{2}-\mu_1)! (\frac{N}{2}+\mu_1)! (\frac{N}{2}-\mu_2)! (\frac{N}{2}+\mu_2)!}, \nonumber
\eeq
where 
\beq \label{AN}
&& \mathcal{A}_N(\mu_2) = 1 + (-1)^{\mu_2} \big[V_N^+ V_N^- + W_N^+ W_N^- \big], \nonumber \\
%\eeq
%\beq \label{CN}
&& \mathcal{C}_N(\mu_1, \mu_2) = (-1)^{\mu_1} \Big[ V_N^+ V_N^- - W_N^+ W_N^- \Big] + (-1)^{\mu_1+\mu_2} 
\bigg(\frac{V_{2N}^+ + V_{2N}^- }{2} \bigg), \nonumber \\
%\eeq
%\beq \label{SN}
&& \mathcal{S}_N(\mu_1, \mu_2) = (-1)^{\mu_1} \Big[ V_N^+ W_N^- + W_N^+ V_N^- \Big] + (-1)^{\mu_1+\mu_2} 
\bigg(\frac{W_{2N}^+ + W_{2N}^- }{2} \bigg), \nonumber \\
\eeq
and 
\beq \label{FourierC}
V_K^{\pm} \equiv \int_{-\pi}^{\pi} d \epsilon_{\pm} P_\pm(\epsilon_\pm) \cos(K \epsilon_{\pm}), 
\quad 
W_K^{\pm} \equiv \int_{-\pi}^{\pi} d \epsilon_{\pm} P_\pm(\epsilon_\pm) \sin(K \epsilon_{\pm}), 
\eeq
$K$ being an integer number.
We are now ready to compute the FI, Eq.~(\ref{Fisher}),
\beq
F(\theta) = \sum_{\mu_1, \mu_2=-N/2}^{N/2} \frac{1}{P(\mu_1, \mu_2 \vert \theta)} \bigg( \frac{d P(\mu_1, \mu_2 \vert \theta)}{d \theta} \bigg)^2.
\nonumber
\eeq
The FI can be written 
as the sum of three terms:
\beq
F(\theta) = N^2 \big[  \mathcal{F}_C(N, \theta) \cos^2 N \theta + 
\mathcal{F}_{SC}(N, \theta) \sin 2 N \theta +
\mathcal{F}_S(N, \theta) \sin^2 N \theta \big], \nonumber \\
\eeq
where the coefficients $\mathcal{F}_C(N, \theta)$, $\mathcal{F}_S(N, \theta)$ 
and $\mathcal{F}_{SC}(N, \theta)$ are function of $N$ and $N\theta$ and are gives by a sums 
over $\mu_{1}$ and $\mu_2$.
To compute the sums we separate the sum over $\mu_{1,2}$ into sum over odd 
$\mu_{1,2}$ and sum over even $\mu_{1,2}$ (since $N$ is assumed to be even, $\mu_{1}$ and $\mu_2$ are integer numbers) 
and take into account that 
\beq
\sum_{\mu, {\rm odd}} \frac{1}{2^N} \frac{N!}{(N/2-\mu)! (N/2+\mu)!} = 
\sum_{\mu, {\rm even}} \frac{1}{2^N} \frac{N!}{(N/2-\mu)! (N/2+\mu)!} = \frac{1}{2}. \nonumber 
\eeq
We thus obtain
\beq
\mathcal{F}_C(N, \theta) &=& \frac{1}{2} \frac{\mathcal{S}_N^2(0, 1) \mathcal{A}_N(1) }
{\mathcal{A}_N^2(1) - [ \mathcal{C}_N (0, 1) \cos N \theta - \mathcal{S}_N(0, 1) \sin N \theta ]^2} + \nonumber \\
&& \quad\quad + \frac{1}{2} \frac{\mathcal{S}_N^2(0, 0) \mathcal{A}_N(0) }
{\mathcal{A}_N^2(0) - [ \mathcal{C}_N (0, 0) \cos N \theta - \mathcal{S}_N(0, 0) \sin N \theta ]^2}, \nonumber
\eeq
\beq
\mathcal{F}_S(N, \theta) &=& \frac{1}{2} \frac{\mathcal{C}_N^2(0, 1) \mathcal{A}_N(1) }
{\mathcal{A}_N^2(1) - [ \mathcal{C}_N (0, 1) \cos N \theta - \mathcal{S}_N(0, 1) \sin N \theta ]^2} + \nonumber \\
&& \quad\quad + \frac{1}{2} \frac{\mathcal{C}_N^2(0, 0) \mathcal{A}_N(0) }
{\mathcal{A}_N^2(0) - [ \mathcal{C}_N (0, 0) \cos N \theta - \mathcal{S}_N(0, 0) \sin N \theta ]^2}, \nonumber
\eeq
and
\beq
\mathcal{F}_{SC}(N, \theta) &=& \frac{1}{2} \frac{ \mathcal{C}_N(0, 1) \mathcal{S}_N(0, 1) \mathcal{A}_N(1) }
{\mathcal{A}_N^2(1) - [ \mathcal{C}_N (0, 1) \cos N \theta - \mathcal{S}_N(0, 1) \sin N \theta ]^2} + \nonumber \\
&& \quad\quad + \frac{1}{2} \frac{ \mathcal{C}_N(0, 0) \mathcal{S}_N(0, 0) \mathcal{A}_N(0) }
{\mathcal{A}_N^2(0) - [ \mathcal{C}_N (0, 0) \cos N \theta - \mathcal{S}_N(0, 0) \sin N \theta ]^2}. \nonumber
\eeq
The above equations allow to calculate the FI given an arbitrary relative and total noise functions.
For the case of NOON input states considered here, the FI ultimately depends on the eight Fourier coefficients 
$V^{\pm}_N$, $V^{\pm}_{2N}$, $W^{\pm}_N$ and $W^{\pm}_{2N}$.
Below, we first shown how the calculation of the FI simplifies when noise distributions are even functions of $\varepsilon_{\pm}$.
Furthermore, we study the case of perfectly correlated relative noise and arbitrary total noise distribution.\\   
\\*
{\it Symmetric noise distributions.} If $P_{\pm}(\epsilon_\pm)$ are even functions of $\epsilon_\pm$, the calculation of the Fsiher information simplify notably. 
We have  $W_K^{\pm}=0$, which implies $\mathcal{S}_N(\mu_1, \mu_2)=0$ for all $\mu_1$ and $\mu_2$, 
$\mathcal{F}_C(N, \theta)=0$ and $\mathcal{F}_{SC}(N, \theta)=0$.
We also have 
$\mathcal{A}_N^2(0) = 1 + V_N^+ V_N^- = A_+$, 
$\mathcal{A}_N^2(1) = 1 - V_N^+ V_N^- = A_-$,
$\mathcal{C}_N (0, 1) = V_N^+ V_N^- - (V_{2N}^+ + V_{2N}^-)/2 = B_-$
and
$\mathcal{C}_N (0, 0) = V_N^+ V_N^- + (V_{2N}^+ + V_{2N}^-)/2 = B_+$,
where $A_{\pm}$ and $B_{\pm}$ have been introduced in Sec.~\ref{Sec.NOON}.
The conditional probability and the FI reduce to Eqs.~(\ref{Probability.NOON}) and 
(\ref{Fisher.NOON}), respectively. \\
\\*
{\it Perfectly correlated relative noise.}
In the following we consider the ideal case of perfectly correlated relative noise, $P_-(\varepsilon_-)=\delta(\varepsilon_-)$.
This implies $V_N^- = V_{2N}^- =1$, $W_N^- = W_{2N}^- =0$ and
Eqs.~(\ref{AN}) simplify to 
\beq
\mathcal{A}_N(\mu_2) &=& 1 + (-1)^{\mu_2}V_N^+,  \nonumber \\
\mathcal{C}_N(\mu_1, \mu_2) &=& (-1)^{\mu_1} V_N^+ + (-1)^{\mu_1+\mu_2} 
\bigg(\frac{1 + V_{2N}^+ }{2} \bigg), \nonumber \\
\mathcal{S}_N(\mu_1, \mu_2) &=& (-1)^{\mu_1}  W_N^+  + (-1)^{\mu_1+\mu_2} \frac{W_{2N}^+}{2}.\nonumber
\eeq
These equation are the basis of further considerations. 
For instance, 
if $P_+(\varepsilon)$ [we indicate $\varepsilon \equiv \varepsilon_+$ to simplify the notation] is an odd function of $\varepsilon$
plus a constant providing normalization in the $2\pi$ interval, then $V_K^+=0$ and we can expand it in Fourier series as 
\beq
P_+(\varepsilon) = \frac{1}{2\pi} + \frac{1}{\pi} \sum_{K=1}^{+\infty} W_K^+ \sin K\varepsilon. \nonumber
\eeq
The condition $P_+(\varepsilon) \geq 0$ implies 
$4 \pi^2  \vert \sum_{K=1}^{+\infty} W_K^+ \sin K\varepsilon \vert^2 \leq 1$ which, integrating over $\varepsilon$
gives $\sum_{K=1}^{+\infty} (W_K^+)^2 \leq 1/2$.
In this case, evaluating $F(\theta)$ at phase values $\theta$ such that $\cos N\theta=0$, we have
[we recall that $F \equiv \max_\theta F(\theta)$]
\beq
\frac{F}{N^2} \geq \frac{1}{8} \bigg( \frac{1}{1- (W_N^+ - W_{2N}^+/2)^2} + \frac{1}{1- (W_N^+ + W_{2N}^+/2)^2} \bigg).
\nonumber
\eeq
The term between brackets does not diverge because of the condition $\sum_{K=1}^{+\infty} (W_K^+)^2 \leq 1/2$
and it is always larger than two. It implies that, in this case $F \geq N^2/4$.
%%%%%%%%%%%%%%%%%%%%%%%%%%%%%%%%%%%%%%%%%%%%%%%%%%%%%%%%%%%%%%%%%%%%%%%%%%%%%%%%%%%%%%%%
\begin{figure}[!t]
\begin{center}
\includegraphics[scale=0.7]{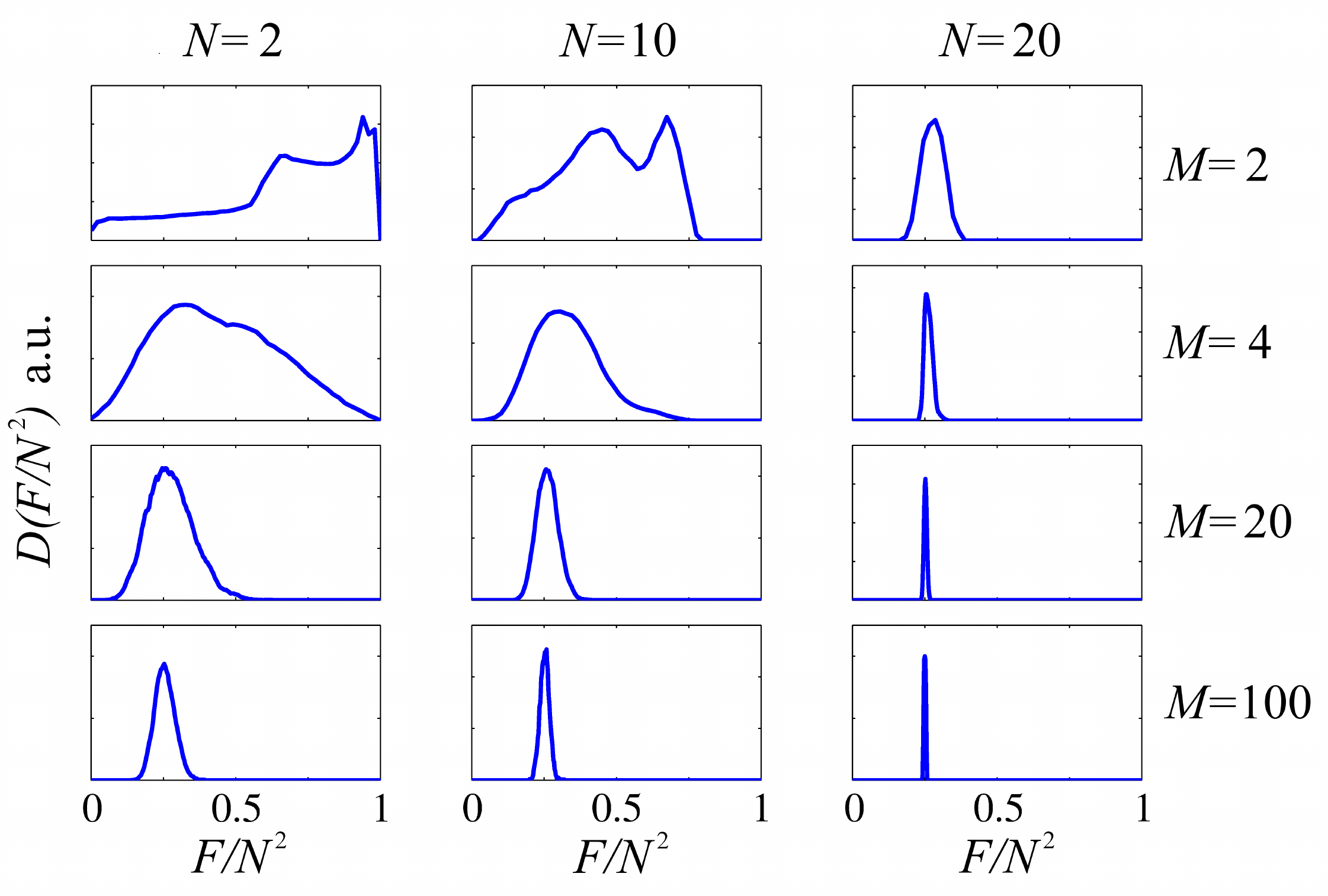}
\end{center}
\caption{\small{ 
Statistical distribution of the $F/N^2$, $D(F/N^2)$, obtained by taking Eq.~(\ref{NoiseMod}) as (total) phase noise distribution. 
The different panels refer to different number of particles, $N$, (different columns) and number, $M$, of noise peaks in Eq.~(\ref{NoiseMod}) (different rows). Here $\sigma=2\pi/100$.
}} \label{Fig:NoiseMod} 
\end{figure} 
%%%%%%%%%%%%%%%%%%%%%%%%%%%%%%%%%%%%%%%%%%%%%%%%%%%%%%%%%%%%%%%%%%%%%%%%%%%%%%%%%%%%%%%%
To treat a more general case, we consider the noise distribution 
\beq \label{NoiseMod}
P_+(\varepsilon) \propto \sum_{n=1}^M e^{\cos(\epsilon-x_n)/\sigma^2}
\eeq
which is a normalized sum of $M$ peaks of width $\sigma$ (for $\sigma \ll 1$ $e^{\cos(\epsilon-x_n)/\sigma^2} \approx e^{-(\epsilon-x_n)^2/2\sigma^2}$, the $\cos$ function being used to take into account the $2\pi$-periodicity) centered at random positions
$x_1,x_2, ... ,x_M \in [-\pi, \pi]$.  
For random choices of $x_1,x_2, ... ,x_M$ we calculate the FI and maximize over $\theta$.
In Fig.~(\ref{Fig:NoiseMod}) we plot the statistical distribution of $F=\max_\theta F(\theta)$ as a function of $N$ and $M$. 

For sufficiently large values of $N$ and/or $M$, the noise distribution (\ref{NoiseMod}) 
has vanishing high frequency Fourier components. 
When increasing $M$ (at fixed value of $N$ and $\sigma$) this is due to the fact that  the noise distribution tends to become 
flat in most of the random realizations (i.e. for most of the random choices of $x_1,x_2, ... ,x_M$).
When increasing $N$ (at fixed $\sigma$ and $\sigma$), this is due to the vanishing tails 
in the Fourier spectrum of $e^{\cos(\epsilon-x_n)/\sigma^2}$.
In both cases, the coefficients $V_N^+$, $W_N^+$, $V_{2N}^+$ and $W_{2N}^-$ are vanishing small, 
and we have
\beq
\mathcal{A}_N(\mu_2)=1, \qquad \mathcal{C}_N(\mu_1, \mu_2) = \frac{(-1)^{\mu_1+\mu_2}}{2}, \qquad \mathcal{S}_N(\mu_1, \mu_2)=0,
\nonumber
\eeq
and 
\beq \label{FthetaN^2/4}
F(\theta) = \frac{N^2 \sin^2 N \theta}{4 - \cos^2 N\theta}, 
\eeq
giving $F \equiv \max_\theta F(\theta)= N^2/4$.
In Fig.~\ref{Fig:NoiseMod} we indeed observe that the distribution of $F$ 
peaks around $1/4$ for sufficiently large values of $N$ and $M$.
  
For small values of $M$ and $N$ we may have a situations where $F/N^2$ is very small. 
In general, for a fixed number of particles, it is possible to derive pathologic noise distributions for 
which the FI vanishes. 
To see this, it is convenient to rewrite $F(\theta)$ as 
\beq
\frac{F(\theta)}{N^2}= \frac{(1-V_N^+)U_1^2(\theta)}{ 2 D_1(\theta)}+\frac{(1+V_N^+)U_0^2(\theta)}{ 2 D_0(\theta)}, 
\eeq
with
\beq
U_j (\theta) &=& \left( W_N^+ + (-1)^j \frac{W_{2N}^+}{2} \right) \cos N\theta + 
\left( V_N^+ + (-1)^j \frac{1+V_{2N}^+}{2} \right) \sin N\theta,  \nonumber
\eeq
and 
\beq 
D_j(\theta)&=& \left(1 + (-1)^j V_N^+\right)^2 -\bigg[ \bigg(V_N^+ + (-1)^j \frac{1+V_{2N}^+}{2} \bigg) \cos N\theta - \nonumber \\
&& \qquad \qquad \qquad \qquad \qquad \qquad 
-\bigg(W_N^+ +(-1)^j \frac{W_{2N}^+}{2} \bigg)\sin N\theta \bigg]^2, \nonumber 
\eeq 
with $j=0,1$.
It's possible to demonstrate that $D_{1,0}>0$ $\forall \theta$. Therefore the Fisher is zero only if both numerators are zero. It is also possible to see that the cases involving $V_N^+=\pm 1$ and $U_{0,1}=0$ lead to non-physical probability distributions. The only remaining option is to have both $U_{0,1}(\theta)=0$ $\forall \theta$. This in turn corresponds to a probability distribution with 
$V_N^+=0$, $W_N^+=0$, $W_{2N}^+=0$ and $V_{2N}^+=-1$.
Recalling the definition of $V_{2N}^+$, we thus have
that $F(\theta)=0$ is and only if
\beq \label{F0cond}
\int d\epsilon \, P(\epsilon)\cos^2 N\epsilon =0.
\eeq
This integral involves two positive functions. Equation~(\ref{F0cond}) is thus fulfilled only if 
$P_+$ to have support in correspondence to the zeroes of $\cos N \epsilon$.
A total noise distribution $P(\epsilon)$ for which the FI vanishes is therefore obtained 
as a normalized sum of Dirac deltas symmetrically centered at the zeroes of $\cos N \epsilon$.
We argue that this situation is pathological for NOON states where the FI is entirely determined 
by the Fourier components of $P(\varepsilon)$, Eq.~(\ref{FourierC}), at $K=N$ and $K=2N$.
Furthermore, if $P(\epsilon)$, instead of being a sum of Dirac peaks, is a sum of peaks of finite width, 
we recover, as noticed above, Eq.~(\ref{FthetaN^2/4}) for $N$ sufficiently large.
}

\section{Numerical Method to compute the Fisher Information}

Here we report a method for the numerical calculation of the FI 
that we used to obtain the results of Sec.~\ref{Sec:BEC}. 
Here we consider a differential interferometer and indicate with 
$\mu_1$ and $\mu_2$ the results of a measurement at the outputs
of the two devices.
The differential interferometer transformation is 
$\hat U(\theta, \varepsilon_1, \varepsilon_2) = e^{-i (\theta + \varepsilon_1) \hat{J}_1} \otimes e^{-i \varepsilon_2 \hat{J}_2}$
and the joint conditional probability reads
 \beq
 P(\mu_1, \mu_2 \vert \theta) = \int_{-\pi}^{\pi} d\epsilon \, P(\mu_1\vert \theta+\epsilon) P(\mu_2\vert \epsilon) P(\epsilon), 
 \eeq
where we have assumed $P_{-}(\epsilon_-)=\delta(\epsilon_-)$.
Noticing that the functions $P(\mu_i|x)$, $i=1,2$, are $2\pi$ periodic in $x$,
it is therefore possible to make a Fourier expansion of the functions.
This is conveniently done with a Fast Fourier Transform algorithm. 
Furthermore, the discretized atom number poses a maximum allowed frequency 
in the decomposition given by Shannon's criterion:
\beq
	P(\mu_i|x)=\frac{1}{2} \sum_{k=-N}^{N} a_k(\mu_i) \cos ( k x ) + b_k(\mu_i) \sin ( k x ), \nonumber
\eeq
where $a_k(\mu_i)$ and $b_k(\mu_i)$ are Fourier coefficients of $P(\mu_i|x)$.
We thus find
\beq \label{Pmumu}
P(\mu_1,\mu_2|\theta) = 
\frac{1}{4} \sum_{k=-N}^{N} A^{(\eta)}_k(\mu_1,\mu_2) \cos ( k \theta ) + B^{(\eta)}_k(\mu_1,\mu_2) \sin ( k \theta ), \nonumber
\eeq
where the coefficients are given by
\beq
A^{(\eta)}_k(\mu_1,\mu_2) = \mathbf{a}^{\rm T}(\mu_1) \mathbf{C} \cdot \mathbf{a}(\mu_2) 
+ \mathbf{b}^{\rm T}(\mu_1) \mathbf{S} \cdot \mathbf{b}(\mu_2)  \nonumber
\eeq
and 
\beq
B^{(\eta)}_k(\mu_1,\mu_2) = \mathbf{b}^{\rm T}(\mu_1) \mathbf{C} \cdot \mathbf{a}(\mu_2)
- \mathbf{a}^{\rm T}(\mu_1) \mathbf{S} \cdot \mathbf{b}(\mu_2),  \nonumber
\eeq
$\mathbf{a}(\mu_i) \equiv (a_{-N}(\mu_i), ..., a_N(\mu_i))$ [and analogous 
definition for $\mathbf{b}(\mu_i)$] are vectors of Fourier coefficients, and 
the matrices $\mathbf{C}$ and $\mathbf{S}$ have components 
\beq
\mathbf{C}_{k,k'}\equiv \int_{0}^{2\pi} d \epsilon \, P(\epsilon)
\cos(k \epsilon) \cos(k' \epsilon),   \quad \,\,
\mathbf{S}_{k,k'} \equiv \int_{0}^{2\pi} d \epsilon \, P(\epsilon)
\cos(k \epsilon) \sin(k' \epsilon),   \nonumber
\eeq
respectively.
Thus, from the knowledge of the Fourier expansion of the conditional probabilities of the single interferometer, 
we can directly find the Fourier expansion of the conditional probability of the differential measurement.
An advantage of this method is that taking the derivative of $P(\mu_1,\mu_2|\theta)$ from Eq.~(\ref{Pmumu}), necessary to calculate the FI, is immediate.

%%%%%%%%%%%%%%%%%%%%%%%%%%%%%%%%%%%%%%%%%%%%%%%%%%%%%%%%%%%%%
%%%%%%%%%%%%%%%%%%%%%%%%%%%%%%%%%%%%%%%%%%%%%%%%%%%%%%%%%%%%%
%%%%%%%%%%%%%%%%%%%%%%%%%%%%%%%%%%%%%%%%%%%%%%%%%%%%%%%%%%%%%

\end{document}